\documentclass[fleqn,twoside,psfig]{article}
\usepackage{espcrc2}

\usepackage{psfig}
\newcommand{\beq}{\begin{equation}}
\newcommand{\eeq}{\end{equation}}
\newcommand{\beqn}{\begin{eqnarray}}
\newcommand{\eeqn}{\end{eqnarray}}
\newcommand{\beqns}{\begin{eqnarray*}}
\newcommand{\eeqns}{\end{eqnarray*}}

\def\PL{{\it Phys. Lett.}}

\def\PRL{{\it Phys. Rev. Lett.}}

\def\ea{{\it et al.}}
\def\Cl{Collaboration}
\def\pc{$\%$}

\def\sfs{spectral functions}

\def\as{$\alpha_s$}
\def\asm{$\alpha_s(m_\tau^2)$}

\def\ee{$e^+e^-$}

\def\FOPTCI{$\rm FOPT_{\rm CI}$}
\def\Rt{$R_\tau$}

\def\RtVpA{$R_{\tau,V+A}$}

\def\RtVpAs{$R_{\tau,V+A}(s_0)$}

\def\piz{\pi^0 }
\newcommand{\phyi}{$e$}
\newcommand{\phyii}{$\mu$}
\newcommand{\phyiii}{$\pi^-$}
\newcommand{\phyiv}{$\pi^-\pi^0$}
\newcommand{\phyv}{$\pi^-2\pi^0$}
\newcommand{\phyvi}{$\pi^-3\pi^0$}
\newcommand{\phyvii}{$\pi^-\pi^-\pi^+$}
\newcommand{\phyviii}{$\pi^-\pi^-\pi^+\pi^0$}
\newcommand{\phyix}{$\pi^-\pi^-\pi^+2\pi^0$}
\newcommand{\phyx}{$\pi^-\pi^-\pi^+3\pi^0$}
\newcommand{\phyxi}{$3\pi^-2\pi^+$}
\newcommand{\phyxii}{$3\pi^-2\pi^+\pi^0$}
\newcommand{\phyxiii}{$\pi^-4\pi^0$}

\def\ie{{\it i.e.}}

\title{Spectral Functions from Hadronic $\tau$ Decays and QCD}

\author{Michel Davier\\
	Laboratoire de l'Acc\'el\'erateur Lin\'eaire\\
        IN2P3/CNRS et Universit\'e de Paris-Sud\\
        91898 Orsay, France\\
	E-mail: davier@lal.in2p3.fr}

\begin{document}

\begin{abstract}
Hadronic decays of the $\tau$ lepton provide a clean environment to
study hadron dynamics in an energy regime dominated by resonances. The
interesting information is captured in the spectral functions. Recent results 
from ALEPH on exclusive channels are presented, with emphasis 
on the $\pi \piz$ final state which plays a crucial role for the determination 
of the hadronic contribution to the muon anomalous magnetic moment.
A comparison between $2\pi$ spectral functions obtained in $\tau$ decays
(after corrections for isospin-breaking) and $e^+e^-$ annihilation reveals
some discrepancy in the line shape of the $\rho$ resonance which
can be attributed to different pole mass values for the charged
and neutral $\rho$'s, which are determined through a robust fitting
procedure. However, after applying this correction, the normalization of the 
two spectral functions differ by $3.3\%$. 
Inclusive spectral functions are the basis for QCD analyses, 
which deliver an accurate determination of the strong
coupling constant, quantitative information on nonperturbative
contributions and a measurement of the mass of the strange quark.
\vspace{1pc}
\end{abstract}

\maketitle

\section{Introduction}

Hadrons produced in $\tau$ decays are born out of the charged weak
current, {\it i.e.} out of the QCD vacuum, which guarantees
that hadronic physics factorizes. These processes are then
completely characterized for each decay channel by a spectral function 
which can be directly extracted from the
invariant mass spectra of the final state. Furthermore, the produced
hadronic systems have $I=1$ and spin-parity $J^P=0^+,1^-$ (V) and
$J^P=0^-,1^+$ (A).  
Isospin symmetry (CVC) connects the $\tau$ and $e^+e^-$ annihilation
spectral functions.

Hadronic $\tau$ decays are a clean probe of hadron dynamics in an
interesting energy region dominated by resonances. However, perturbative
QCD can be seriously considered due to the relatively large $\tau$ mass.
Samples of $\sim 4 \times 10^5$ measured decays
are available in each LEP experiment and CLEO. Conditions for low 
systematic uncertainties are particularly well met at LEP: measured
samples have small non-$\tau$ backgrounds ($\sim 1\%$) and large selection
efficiency ($92\%$), for example in ALEPH.

\section{Spectral Functions from ALEPH}

Preliminary spectral functions based on the full LEP1 statistics are 
available from ALEPH. The corresponding results~\cite{tau02_br} 
for the branching fractions
which provide the absolute normalization for the spectral functions 
are given in Table~\ref{table_br}. The analysis uses an improved treatment
of photons as compared to the published analyses based on a reduced 
sample~\cite{aleph_v,aleph_a}. Spectral functions are unfolded from the
measured mass spectra after background subtraction using a mass-migration 
matrix obtained from the simulation in order to account for detector 
and reconstruction biases. Backgrounds from non-$\tau \tau$ events 
are small ($< 1 \%$ and subtractions are dominated by $\tau$ decay 
feedthroughs. 

\begin{table}
\begin{center}
\begin{tabular}{lrr}
\hline\hline
 mode & B $\pm\sigma_{\hbox{tot}}$
   [\%] & {\footnotesize  ALEPH Prel.}\\\hline
   \phyi &    17.837 $\pm$  0.080 &\\
   \phyii &    17.319 $\pm$  0.077 &\\
\hline
   \phyiii &    10.828 $\pm$  0.105 &A\\
   \phyiv &    25.471 $\pm$  0.129  &V\\
   \phyv &     9.239 $\pm$  0.124  &A\\
   \phyvi &      0.977 $\pm$  0.090  &V\\
   \phyxiii &      0.112 $\pm$  0.051  &A\\
   \phyvii &     9.041 $\pm$  0.097  &A\\
   \phyviii &     4.590 $\pm$  0.086  &V\\
   \phyix &      0.392 $\pm$  0.046  &A\\
   \phyx &      0.013 $\pm$  0.010  &V\\
   \phyxi &      0.072 $\pm$  0.015  &A\\
   \phyxii &      0.014 $\pm$  0.009  &V\\
   $\pi^- \pi^0 \eta$ & 0.180 $\pm$ 0.045  &V\\
   $(3\pi)^-  \eta$ & 0.039 $\pm$ 0.007  &A\\
   $a_1^- (\rightarrow \pi^- \gamma)$ & 0.040 $\pm$ 0.020  &A\\
   $\pi^- \omega (*)$ & 0.253 $\pm$ 0.018     &V\\
   $\pi^- \pi^0 \omega (*)$ & 0.048 $\pm$
   0.009  &A\\
   $(3\pi)^- \omega (*)$ & 0.003 $\pm$
   0.003  &V\\
\hline
   $K^- K^0 $ &  0.163 $\pm$ 0.027 &V\\
   $K^- \pi^0 K^0 $    &  0.145 $\pm$ 0.027  & $(94^{+6}_{-8})\%$A\\
   $\pi^- K^0 \overline{K^0}$
                &  0.153 $\pm$ 0.035  &$(94^{+6}_{-8})\%$A\\
   $K^- K^+ \pi^- $    &  0.163 $\pm$ 0.027  &$(94^{+6}_{-8})\%$A\\
   $(K\overline{K}\pi\pi)^-$ &  0.050  $\pm$ 0.020   &$(50 \pm 50)\%$ A\\
\hline
   $K^- $ &  0.696 $\pm$ 0.029 &S\\
   $K^- \pi^0 $    &  0.444 $\pm$ 0.035  &S\\
   $\overline{K^0} \pi^-$ &  0.917 $\pm$ 0.052  &S\\
   $K^- 2\pi^0 $    &  0.056 $\pm$ 0.025  &S\\ 
   $K^- \pi^+ \pi^- $    &  0.214 $\pm$ 0.047  &S\\
   $\overline{K^0} \pi^- \piz$ &  0.327 $\pm$ 0.051  &S\\
   $(K 3\pi)^- $ &  0.076 $\pm$ 0.044 &S\\
   $K^- \eta$  &  0.029 $\pm$ 0.014 &S\\
\hline\hline
\end{tabular}
\caption{Branching fractions in $\tau$ decays from the ALEPH 
experiment~\cite{tau02_br,aleph_ksum}. Apart from the two leptonic channels,
the other modes are labeled according to their nonstrange vector (V), 
nonstrange axial-vector (A), and strange (S) hadronic final states. The
$\omega$ decay modes marked (*) are electromagnetic ($\pi^0 \gamma$, 
$\pi^+ \pi^-$). The branching fractions for \phyx\ and 
$a_1^- (\rightarrow \pi^- \gamma)$ are estimates,while those for
$(3\pi)^-  \eta$ and $(3\pi)^- \omega$ are from CLEO~\cite{cleo_3piX}.}
\label{table_br}
\end{center}
\end{table}

The spectral functions are separated into vector and axial-vector components
according to the number of pions in the final state. 
As for final states involving $K\overline{K}$ pairs, specific input 
is required to achieve the $V-A$ separation~\cite{aleph_ksum}.

\section{The $2\pi$ Vector State}

\subsection{The Data}

\begin{figure}[t]
   \centerline{\psfig{file=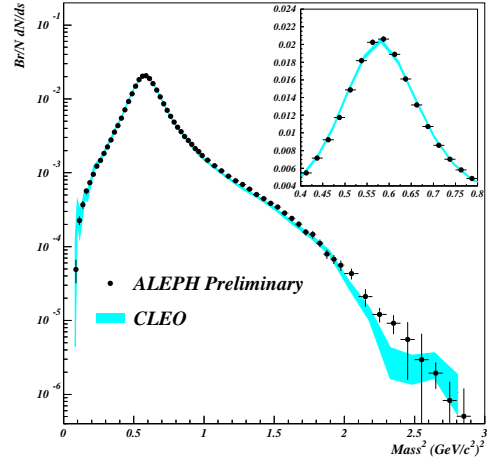,width=70mm}}
   \caption{The comparison between ALEPH and CLEO spectral functions
in the $\pi \pi^0$ mode. }
\label{aleph_cleo_2pi}
\end{figure}

The spectral function from $\tau \rightarrow \nu_\tau \pi^- \pi^0$ 
from the full-LEP1 ALEPH analysis ($\sim 10^5$ events) is available.
It is in good agreement with the results from CLEO~\cite{cleo_2pi}
as shown in Fig.~\ref{aleph_cleo_2pi}. 
The statistics is comparable in the two cases, however due to a flat
acceptance in ALEPH and a strongly increasing one in CLEO, ALEPH
data are more precise below the $\rho$ peak, while CLEO is more precise
above. Note that, due to the unfolding procedure, neighbouring data points
are strongly correlated.

\subsection{$\pi \pi$ Spectral Functions and $\pi$ Form Factors}

It is useful to carefully write down all the factors involved in the 
comparison of \ee\ and $\tau$ spectral functions in order to make
explicit the possible sources of CVC breaking. On the \ee\ side we have
\begin{eqnarray}
\sigma (e^+e^-\longrightarrow \pi^+\pi^-)&=&\frac{4\pi\alpha^2}{s} v_0(s)\\
    v_0(s)&=&\frac {\beta_0^3(s)} {12 \pi} |F^0_\pi(s)|^2 \nonumber
\label{ee_ff}
\end{eqnarray}
where $\beta_0^3(s)$ is the threshold kinematic factor and $F^0_\pi(s)$
the pion form factor. On the $\tau$ side, the physics is contained in the
hadronic mass distribution through 
\begin{eqnarray}
\frac {1}{\Gamma} \frac {d \Gamma}{ds}
 (\tau \longrightarrow \pi^-\pi^0 \nu_\tau) &=& \nonumber \\
         & & \hspace{-2.7cm}
                    \frac {6 \pi |V_{ud}|^2 S_{EW}}{m_\tau^2}
                    \frac {B_e}{B_{\pi \pi^0}} C(s) v_-(s) \\
      v_-(s) &=& \frac {\beta_-^3(s)} {12 \pi} |F^-_\pi(s)|^2 \nonumber \\
      C(s) &=& \left(1- \frac {s}{m_\tau^2}\right) \nonumber
       \left(1 + \frac {2 s}{m_\tau^2}\right)
\label{tau_ff}
\end{eqnarray}
where $V_{ud}=0.9748 \pm 0.0010$ denotes the CKM 
weak mixing matrix element~\cite{pdg2002} and $S_{\mathrm{EW}}$ 
accounts for electroweak radiative corrections (see below).
SU(2) symmetry implies $v_-(s) =  v_0(s)$.

Experiments on $\tau$ decays measure the rate
inclusive of radiative photons, {\it i.e.} for 
$\tau \rightarrow \nu_\tau \pi \piz (\gamma)$. The measured spectral function
is thus $v_-^*(s) =  v_-(s)~G(s)$, where $G(s)$ is a radiative correction,
computed using scalar QED.

Several levels of SU(2) breaking can be identified:
\begin{itemize}
\item {\it electroweak radiative corrections to $\tau$ decays} 
are contained in the $S_{\mathrm{EW}}$ 
factor~\cite{marciano,braaten} which is dominated by short-distance effects.
As such it is expected to be weakly dependent on the specific hadronic
final state, as verified in the 
$\tau \longrightarrow (\pi, K) \nu_\tau$ channels~\cite{decker}. 
Recently, detailed calculations have been performed for the $\pi \pi^0$
channel~\cite{cen} which also confirm the relative smallness of the
long-distance contributions. The total correction is
\beq
 S_{EW} = \frac {S_{EW}^{\rm had} S_{EM}^{\rm had} }{S_{EM}^{\rm lep}}
\eeq
where $S_{EW}^{\rm had}$ is the leading-log short-distance electroweak
factor (which vanishes for leptons) and $S_{EM}^{\rm had,lep}$ are the
nonleading electromagnetic corrections. The latter corrections are
calculated in Ref.~\cite{braaten} at the quark level and in
Ref.~\cite{cen} at the hadron level for the $\pi \pi^0$ decay mode,
and in Ref.~\cite{marciano,braaten} for leptons. The total correction
amounts~\cite{dehz} to $S_{EW}^{\rm inclu} = 1.0198 \pm 0.0006$ 
for the inclusive hadron decay rate and 
$S_{EW}^{\pi \pi^0} = (1.0232 \pm 0.0006)~S_{EM}^{\pi \pi^0}(s)$ 
for the $\pi \pi^0$ decay mode, where $S_{EM}^{\pi \pi^0}(s)$ is an
$s$-dependent radiative correction~\cite{cen}.
\item {\it the pion mass splitting} breaks isospin symmetry in 
the spectral functions~\cite{adh,czyz} since $\beta_-(s) \neq \beta_0(s)$.
\item symmetry is also broken in {\it the pion form factor}
~\cite{adh,cen} from the $\pi$ mass splitting.
\item a similar effect is expected from {\it the $\rho$ mass splitting}. 
The theoretical expectation~\cite{bijnens} gives a limit ($<0.7$~MeV), 
but it is really only a rough estimate. Thus the question must be investigated
experimentally, the best approach being the explicit comparison
of $\tau$ and $e^+e^-$ $2\pi$ spectral functions, after correction for 
the other isospin-breaking effects. 
\item explicit {\it electromagnetic decays} such as 
$\pi \gamma$, $\eta \gamma$, $l^+l^-$ and $\pi \pi \gamma$ introduce some
small difference between the widths of the charged and neutral $\rho$'s.
\item isospin violation in the strong amplitude through the {\it mass 
difference between u and d quarks} is expected to be negligible. 
\end{itemize}

\subsection{Fitting the $2\pi$ Spectral Functions: 
the $\rho$ mass splitting}

The $2\pi$ spectral function is dominated by the wide $\rho$ resonance, 
parametrized following Gounaris-Sakurai~\cite{gounaris} (GS) which takes
into account analyticity and unitarity properties. 

The pion form factor is fit with interfering amplitudes from $\rho(770)$, 
$\rho'(1450)$ and $\rho''(1700)$ vector mesons with relative strengths 1, 
$\beta$ and $\gamma$ (real numbers). A phase $\phi_\beta$ is also considered, 
since the relative phase of the $\rho$ and $\rho'$ amplitude is a priori 
unknown. The much smaller relative amplitude for the $\rho''$ is assumed 
to be  real. In practice we fit $F_\pi^0(s)$ from 
$e^+e^-$ data and $F_\pi^-(s)$ from the $\tau$ spectral 
function duly corrected for SU(2) breaking, however only for the spectral 
function $\beta^3$ factor, for the $S_{EW}$ factor, and for the long-distance 
radiative correction $G_{EM}(s)$. In this way, the mass and width 
of the dominating $\rho$ resonance in the two isospin states can be 
unambiguously determined.
All \ee\ data are used, including the 
recently corrected precise results from CMD-2~\cite{cmd2,dehz03}. 
On the $\tau$ side, the accurate data from ALEPH and CLEO are used.

The systematic uncertainties are included in the fits through appropriate
covariance matrices. The $\rho$ mass systematic uncertainty in $\tau$ data
is mostly from calibration (0.7 MeV for ALEPH and 0.9 MeV for CLEO). The
corresponding uncertainties on the $\rho$ width are 0.8 and 0.7 MeV. Both
mass and width determinations are limited by systematic uncertainties in 
$\tau$ data, however they can be safely assumed to be uncorrelated 
between ALEPH and CLEO.
 
Due to the large event statistics, the fits are quite sensitive to the 
precise line shape and to the interference between the different amplitudes. 
Of course, $\rho-\omega$ interference is included for \ee\ data only and the
corresponding amplitude ($\alpha_{\rho \omega}$) is fit, together with its
relative phase. 
%Fits are performed separately for the $\tau$ and \ee\ form factors. 
The upper range of the fit is taken at 2.4 GeV$^2$ in the $\tau$ data 
in order to avoid the kinematic end-point of the $\tau$ spectral 
function where large corrections are applied. The \ee\ data are fit up to
3.6 GeV$^2$, so that the information on the $\rho''$ comes essentially
from \ee\ data.

It is customary in the Novosibirsk analyses to define $F_\pi$
including the vacuum polarization (both leptonic and hadronic) in the 
photon propagator. Such a prescription is not desirable when studying the
real hadronic structure of the pion, and incorrect when comparing to the
$\tau$ case where these contributions are absent. Removing 
vacuum polarization  in the fit to CMD-2 $|F_\pi|^2$ data
(Table~\ref{fpi_vp-or-not-vp})  yields a $\rho^0$ mass 1.1~MeV lower. 
Subsequent fits are done excluding vacuum polarization.

\begin{table}
\begin{center}
{
%\small
\begin{tabular}{|c|c|c|} \hline 
           &  VP removed  & VP included   \\
\hline \hline
$m_{\rho}$      & 774.4 $\pm$ 0.6  & 775.5 $\pm$ 0.6   \\
$\Gamma_{\rho}$ & 146.7 $\pm$ 1.3  & 145.4 $\pm$ 1.3   \\
%$m_{\rho}$      & 775.6 $\pm$ 0.4  & 775.5 $\pm$ 0.4    \\
%$\Gamma_{\rho}$ & 148.8 $\pm$ 0.8  & 150.3 $\pm$ 0.9    \\
\hline
$\alpha_{\rho\omega}$& $(1.83 \pm 0.14)~10^{-3}$&$(1.51 \pm 0.14)~10^{-3}$  \\
\hline
$\beta$   & 0.079 $\pm$ 0.006 & 0.084 $\pm$ 0.006 \\
$\phi_\beta$   & [180] & [180]   \\
$m_{\rho'}$   & [1465]  & [1465]   \\
$\Gamma_{\rho'}$  & [310]          &  [310] \\   
\hline
$\gamma$   &   [0]   & [0]    \\
%$\phi_\gamma$   & 0. & 0. \\
%$m_{\rho''}$      &  1700    & 1700      \\
%$\Gamma_{\rho''}$   &  235    &  235     \\
%\hline
%$|F^0_\pi (0)|^2$  &  1.  & 0.926 $\pm$ 0.015   \\ 
%$|F^-_\pi (0)|^2$  &  1.  & 1.003 $\pm$ 0.013    \\ 
\hline\hline
$\chi^2$/DF             & 38/39  & 34/39     \\
\hline
\hline
\end{tabular}
}
\caption{Fits to CMD-2 $|F_\pi|^2$ data~\cite{cmd2} using the Gounaris-Sakurai
parametrization of the $\rho$, $\rho'$ resonances for two definitions 
of the pion form factor: excluding or including the vacuum polarization in 
the photon propagator. A mass shift of 1.1~MeV is observed for the $\rho$ 
mass between the two cases. The values between squared brackets are fixed 
in the fits. The masses and widths are in MeV and $\phi_\beta$ in degrees.
%Errors are not repeated for the fit including vacuum polarization.
}
\label{fpi_vp-or-not-vp}
\end{center}
\end{table}

It turns out that the resulting $\rho$ masses and widths are quite sensitive 
to the strength of the $\rho'$ and $\rho''$ amplitudes, $\beta$ and $\gamma$. 
So, depending on the type of fit, the derived values can exhibit some 
systematic shifts. The \ee\ and $\tau$ fits yield significantly
different $\rho'$ amplitudes and phases: this is particularly true when
a restricted energy range is used for the fit, as it is the case when only
CMD-2 data are considered (only up to 960~MeV).
In order to avoid this problem, a fit to both data sets is performed, 
keeping common values for the $\rho'$ and $\rho''$ parameters. In doing so,
one neglects possible isospin-breaking effects for these states, which 
appears to be a reasonable assumption, as the $\rho'$ and $\rho''$
components can be considered as second-order with respect 
to the dominant investigated $\rho$ amplitude.

The differences between the masses and widths of the charged 
and neutral $\rho$'s in the common fit are found to be
\begin{eqnarray}
   m_{\rho^-}-m_{\rho^0}~&=&~ (2.3 \pm 0.8)~{\rm MeV} \\
   \Gamma_{\rho^-}-\Gamma_{\rho^0}~&=&~ (0.1 \pm 1.4)~{\rm MeV}
\label{splitting}
\end{eqnarray}
The mass splitting is somewhat larger than the theoretical prediction 
($<0.7$ MeV)~\cite{bijnens}, but only at the $2~\sigma$ level. The expected
width splitting, not taking into account any $\rho$ mass splitting, is 
$(0.7\pm0.3)$ MeV~\cite{cen,dehz}. However,
if the mass difference is taken as an experimental fact, 
then a larger width difference would be expected. From
the chiral model of the $\rho$ resonance~\cite{pich-portoles,cen},
one expects
\beq
\Gamma_{\rho^0}~=~\Gamma_{\rho^-}\left(\frac {m_{\rho^0}}{m_{\rho^-}}\right)^3 
              \left(\frac {\beta_0}{\beta_-}\right)^3~+~\Delta \Gamma_{EM}
\label{gamma_calc}
\eeq  
where $\Delta \Gamma_{EM}$ is the width difference from electromagnetic decays
(as discussed above),
leading to a total width difference ($2.1 \pm 0.5$) MeV, marginally 
consistent with the observed value.

\begin{figure}[t]
   \centerline{\psfig{file=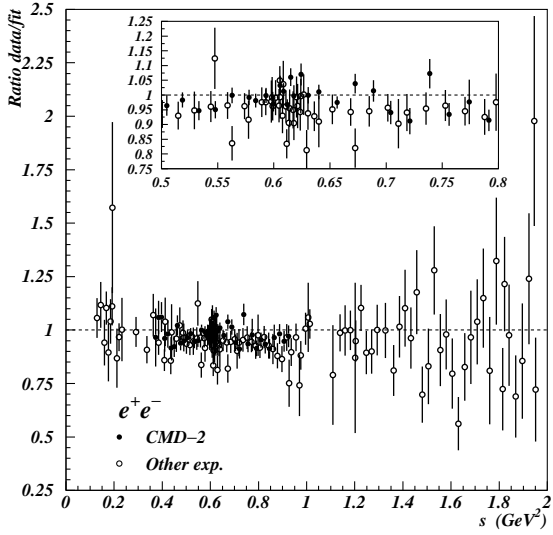,width=80mm}}
   \caption{The ratio of the \ee\ pion form factor squared to 
   $|F_\pi^0(s)|^2$ from the combined fit in Table~\ref{2pi_fits}.}
\label{fit_ee_2pi}
\end{figure}

\begin{figure}[t]
   \centerline{\psfig{file=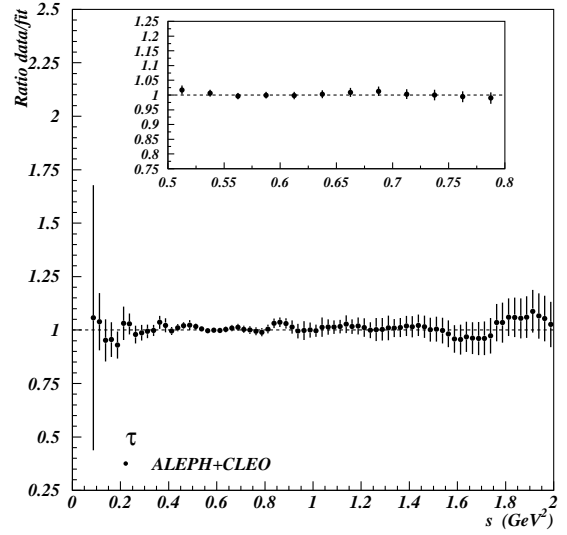,width=80mm}}
   \caption{The ratio of the $\tau$ pion form factor squared to 
   $|F_\pi^-(s)|^2$ from the combined fit in Table~\ref{2pi_fits}. }
\label{fit_tau_2pi}
\end{figure}

Table~\ref{2pi_fits} presents the results of the common fits, the quality
of which can be inspected in Figs.~\ref{fit_ee_2pi} and \ref{fit_tau_2pi}. 

Mass splitting for the $\rho$ was in fact considered in our first paper where 
we proposed using precise $\tau$ data to compute hadronic vacuum polarization
integrals~\cite{adh}. Using pre-CMD-2 \ee\ data and ALEPH $\tau$, a combined 
fit was attempted which produced a mass splitting consitent with 0 within an
uncertainty of 1.1 MeV. However, the form factor from \ee\ data still 
contained the vacuum polarization contribution (1.1 MeV shift, as we have 
seen) and we also discovered a normalization problem in our treatment 
of the $\tau$ data in the combined fit. 
With the advent of precise CMD-2 data~\cite{cmd2}, it became apparent that
differences were showing up between $\tau$ and \ee\ form factors. A large
part of the discrepancy was removed when CMD-2 re-analyzed their 
data~\cite{cmd2}. Since the $\tau$ results from ALEPH, CLEO and OPAL have 
been shown to be consistent within their quoted accuracy~\cite{dehz,dehz03} 
and since preliminary
results from the radiative return analysis of KLOE~\cite{kloe_pisa} are 
in excellent agreement with the corrected CMD-2 results, the question of the
$\rho$ mass splitting can be now more reliably investigated.
Table~\ref{2pi_fits} presents the results of the common fit, leaving
$m_{\rho^-}$ and $m_{\rho^0}$ as free parameters, but fixing the
relationship between $\Gamma_{\rho^-}$ and $\Gamma_{\rho^0}$ following
Eq.~(\ref{gamma_calc}). The results, which can be visualized 
in Figs.~\ref{fit_ee_2pi} and \ref{fit_tau_2pi}, do not provide a good
description of the data with a probability of only $0.6\%$ to get a worse fit. 

Could a $\rho$ mass splitting account for the discrepancy~\cite{dehz,dehz03} 
between \ee\ and isospin-corrected $\tau$ spectral functions? Unfortunately
not! Correcting the $\tau$ data for the mass shift using the Gounaris-Sakurai
parametrization with $\rho$, $\rho'$ and $\rho''$ amplitudes (but the results
only depend on a Breit-Wigner-like behaviour for the $\rho$) improves the
consistency of the \ee\ and $\tau$ line shapes, but at the expense of an
increased overall normalization difference: the corrected $\tau$ data 
lie on average $3.3\%$ above \ee\, hence the bad fit. The tau estimate of 
$a_\mu^{\rm had,LO}$ increases by $5.4~10^{-10}$~\cite{md_g-2_pisa}, 
in larger disagreement with the \ee\ estimate~\cite{dehz03}.

\begin{table}
\begin{center}
{\small
\begin{tabular}{|c|c|} \hline 
           &  $\tau$  and \ee\   \\
\hline \hline
$m_{\rho^-}$      & 775.4 $\pm$ 0.6  \\
$m_{\rho^0}$      & 773.1 $\pm$ 0.5  \\
$\Gamma_{\rho^-}$ & 148.8 $\pm$ 0.8  \\
$\Gamma_{\rho^0}$ & (146.7)          \\
\hline
$\alpha_{\rho\omega}$ & $ (2.02 \pm 0.10)~10^{-3}$  \\
$\phi_\alpha$         & $ (15.3 \pm 2.0)$  \\
\hline
$\beta$   & 0.167 $\pm$ 0.006  \\
$\phi_\beta$   & 177.5 $\pm$ 6.0    \\
$m_{\rho'}$   & 1410 $\pm$ 16    \\
$\Gamma_{\rho'}$  & 505 $\pm$ 53          \\   
\hline
$\gamma$   &   0.071 $\pm$ 0.007       \\
$\phi_\gamma$   & 0. \\
$m_{\rho''}$      &  1748 $\pm$ 21         \\
$\Gamma_{\rho''}$   &  235       \\
%\hline
%$|F^0_\pi (0)|^2$  &  1.  & 0.926 $\pm$ 0.015   \\ 
%$|F^-_\pi (0)|^2$  &  1.  & 1.003 $\pm$ 0.013    \\ 
\hline\hline
$\chi^2$/DF             & 394./327      \\
\hline
\hline
\end{tabular}
}
\caption{
         Combined fit to the pion form factor squared to $\tau$ and \ee\ 
         data, vacuum polarization excluded for the latter. 
         The parametrization of the $\rho$, $\rho'$, $\rho''$ line shapes 
         follows the Gounaris-Sakurai form. For the $\rho$, only 
         $m_{\rho^-}$, $m_{\rho^0}$, and $\Gamma_{\rho^-}$ are fitted, while
         $\Gamma_{\rho^0}$ is computed from Eq.~(\ref{gamma_calc}).
         All mass and width values are in MeV and the phases
         are in degrees. The parameters related to $\rho'$ and 
         $\rho''$ amplitudes are fitted, assuming they are identical in
         both data sets.}
\label{2pi_fits}
\end{center}
\end{table}

\section{Inclusive Nonstrange Spectral Functions}

\begin{figure}[t]
   \centerline{\psfig{file=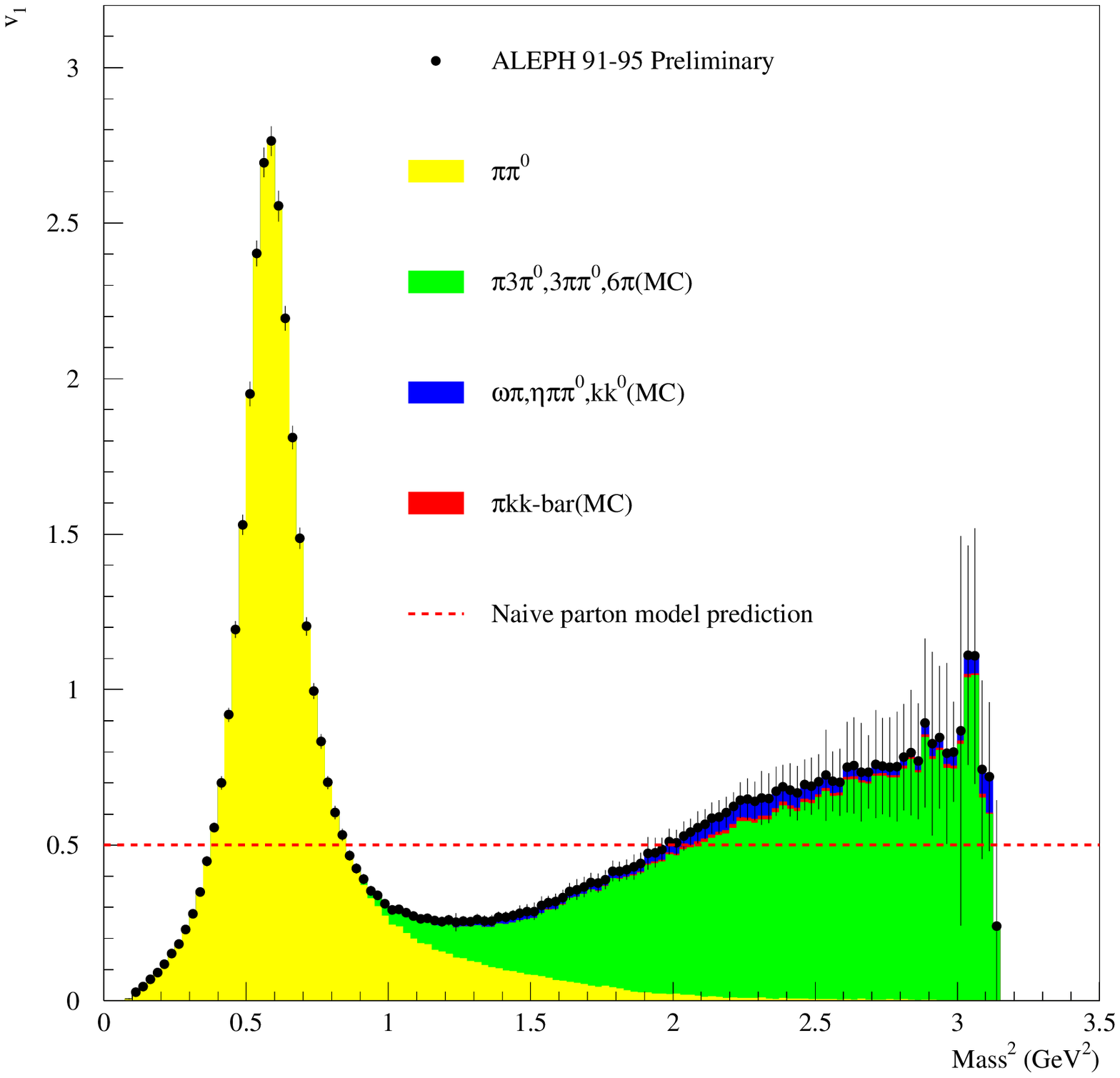,width=70mm}}
   \caption{Preliminary ALEPH inclusive vector spectral function with its
    different contributions. The dashed line is the expectation from 
    the naive parton model.}
\label{v_aleph}
\end{figure}

\begin{figure}[t]
   \centerline{\psfig{file=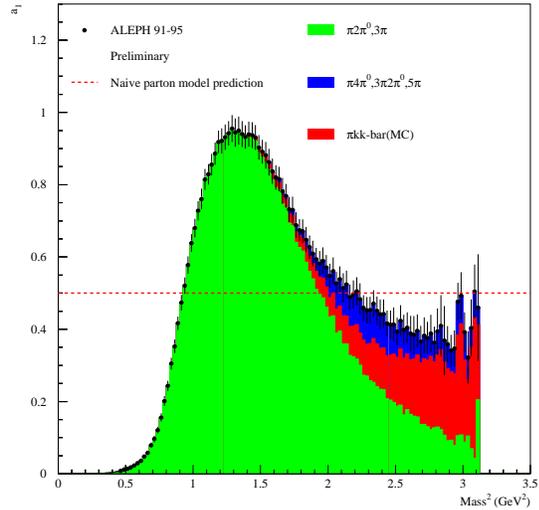,width=70mm}}
   \caption{Preliminary ALEPH inclusive axial-vector spectral function with its
    different contributions. The dashed line is the expectation from 
    the naive parton model. }
\label{a_aleph}
\end{figure}

\begin{figure}[t]
   \centerline{\psfig{file=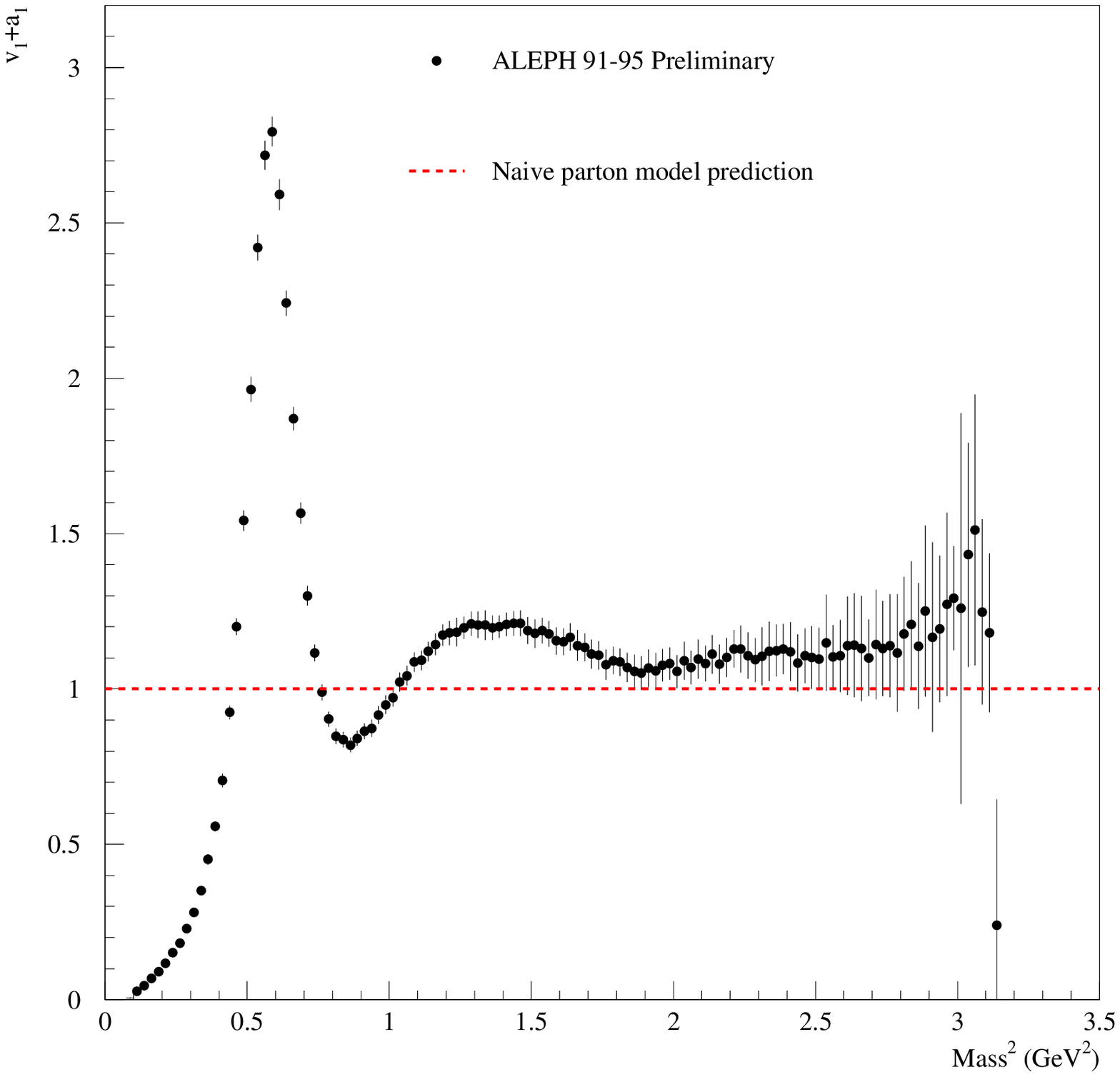,width=70mm}}
   \caption{The preliminary inclusive $V+A$ nonstrange spectral function 
      from ALEPH. The dashed line is the expectation from the naive 
      parton model.}
\label{vpa_aleph}
\end{figure}

The $\tau$ nonstrange spectral functions have been measured by ALEPH
~\cite{aleph_v,aleph_a} and OPAL~\cite{opal}. The procedure requires a
careful separation of vector (V) and axial-vector (A) states involving
the reconstruction of multi-$\pi^0$ decays and the proper treatment of
final states with a $K \bar K$ pair. 
The inclusive ALEPH $V$ and $A$ spectral functions are given in 
Figs.~\ref{v_aleph} and \ref{a_aleph} with a breakdown of the respective
contributions.
The $V+A$ spectral function, shown in Fig.~\ref{vpa_aleph} has a clear 
pattern converging toward a value above the parton level as expected in
QCD. In fact, it displays a textbook example of global duality, since
the resonance-dominated low-mass region shows an oscillatory behaviour around
the asymptotic QCD expectation, assumed to be valid in a local sense only
for large masses. This feature will be quantitatively discussed in the 
next section.

\section{QCD Analysis of Nonstrange $\tau$ Decays}

\subsection{Motivation}

The total hadronic $\tau$ width, properly normalized to the known leptonic
width,

\beq
     R_\tau = \frac{\Gamma(\tau^-\rightarrow{\rm hadrons}^-\,\nu_\tau)}
                   {\Gamma(\tau^-\rightarrow e^-\,\bar{\nu}_e\nu_\tau)}
\eeq
should be well predicted by QCD as it is an inclusive observable. Compared
to the similar quantity defined in \ee\ annihilation, it is even twice
inclusive: not only are all produced hadronic states at a given mass summed
over, but an integration is performed over all the possible masses from
$m_{\pi}$ to $m_{\tau}$.

This favourable situation could be spoiled by the fact that the $Q^2$ scale
is rather small, so that questions about the validity of a perturbative
approach can be raised. At least two levels are to be considered: the 
convergence of the perturbative expansion and the control of the
nonperturbative contributions. Happy circumstances make
these latter components indeed very small~\cite{braaten88,narpic88,bnp}.

The perturbative expansion (FOPT) is known to third order~\cite{3loop}.
A resummation of all known higher order 
logarithmic integrals improves the convergence 
of the perturbative series (contour-improved method \FOPTCI)~\cite{pert}. 
As some ambiguity persists, the results are given as an average of the
two methods with the difference taken as a systematic uncertainty. 
The small nonperturbative contributions are parametrized using the
Operator Product Expansion (OPE)~\cite{svz}.

\subsection{Results of the Fits}

The QCD analysis of the $\tau$ hadronic width has not yet been completed 
with the final ALEPH spectral functions. Results given below correspond
to the published analyses with a smaller data set.

The ratio \Rt\ is obtained from measurements of the leptonic branching 
ratios:
\beqn
  R_\tau&=&3.647\pm0.014 
\eeqn
using a value 
$B(\tau^-\rightarrow e^-\,\bar{\nu}_e\nu_\tau)=(17.794\pm0.045)\%$
which includes the improvement in accuracy provided by the 
universality assumption of leptonic currents together with the 
measurements of $B(\tau^-\rightarrow e^-\,\bar{\nu}_e\nu_\tau)$,
$B(\tau^-\rightarrow \mu^-\,\bar{\nu}_\mu\nu_\tau)$ and the $\tau$ 
lifetime. The nonstrange part of \Rt\ is obtained by subtracting
out the measured strange contribution (see last section).
The results of the fits are given in Table~\ref{tab_asresults} for
the ALEPH analysis.
\begin{table}
{\small
  \begin{tabular}{|l||c|c|} \hline 
     & &  \\
 ALEPH&$\alpha_s(m_{\tau}^2)$       &$\delta_{\rm NP}$\\
\hline 
 V    &  $0.330\pm0.014\pm0.018$    &  $0.020\pm0.004$  \\
 A    &  $0.339\pm0.013\pm0.018$    &  $-0.027\pm0.004$  \\
\hline
 V+A  &  $0.334\pm0.007\pm0.021$    &  $-0.003\pm0.004$  \\
\hline
  \end{tabular}
}
  \caption{
              F\/it results of \asm\ and the OPE nonperturbative 
              contributions from vector, axial-vector and $(V+A)$ combined
              fits using the corresponding ratios \Rt\ and the spectral 
              moments as input parameters. The second error is given for 
              theoretical uncertainty.}
\label{tab_asresults}
\end{table}

One notices a remarkable agreement within statistical errors
between the \asm\ values using vector and axial-vector data.
The total nonperturbative power contribution to \RtVpA\ is compatible 
with zero within an uncertainty of 0.4\pc, that is much smaller than 
the error arising from the perturbative term. This cancellation of the 
nonperturbative terms increases the confidence on the \asm\ 
determination from the inclusive $(V+A)$ observables.

The f\/inal result from ALEPH is : 
\beq
\label{eq_asres}
   \alpha_s(m_\tau^2)    = 0.334 \pm 0.007_{\rm exp} 
                                 \pm 0.021_{\rm th} 
\eeq
where the f\/irst error accounts for the experimental uncertainty and 
the second gives the uncertainty of the theoretical prediction of
$R_\tau$ and the spectral moments as well as the ambiguity of the 
theoretical approaches employed.

\subsection{Test of the Running of \boldmath$\alpha_s(s)$ at Low Energies}

Using the \sfs, one can simulate the physics of a hypothetical 
$\tau$ lepton with a mass $\sqrt{s_0}$ smaller than $m_\tau$
and hence further investigate QCD
phenomena at low energies. Assuming quark-hadron duality, 
the evolution of $R_\tau(s_0)$ provides a direct test of the 
running of $\alpha_s(s_0)$, governed by the RGE $\beta$-function. 
On the other hand, it is a test of the validity of the OPE approach 
in $\tau$ decays. 

In practice, the experimental value for $\alpha_s(s_0)$ 
has been determined at every $s_0$ from the comparison 
of data and theory. Good agreement is observed with the four-loop 
RGE evolution using three quark f\/lavours (Fig.~\ref{fig_runas}).
The experimental fact that the nonperturbative contributions 
cancel over the whole range $1.2~{\rm GeV}^2\le s_0\le m_\tau^2$ 
leads to conf\/idence that the \as\ determination from the inclusive 
$(V+A)$ data is robust.  

\subsection{Evolving $\alpha_s$ from $m_\tau)$ to $M_{\rm Z}$}

The evolution of the 
\asm\ measurement from the inclusive $(V+A)$ observables based on 
the Runge-Kutta integration of the dif\/ferential equation 
of the renormalization group to 
N$^3$LO~\cite{alpha_evol,pichsanta} yields for the ALEPH analysis
\beq
\label{alphaevol}  
   \alpha_s(M_{\rm Z}^2) = \nonumber \\
                           0.1202 \pm 0.0008_{\rm exp} 
                                  \pm 0.0024_{\rm th} 
                                  \pm 0.0010_{\rm evol}
\eeq
where the last error stands for
possible ambiguities in the evolution due to uncertainties in the 
matching scales of the quark thresholds~\cite{pichsanta}. 
\begin{figure}
        \psfig{file=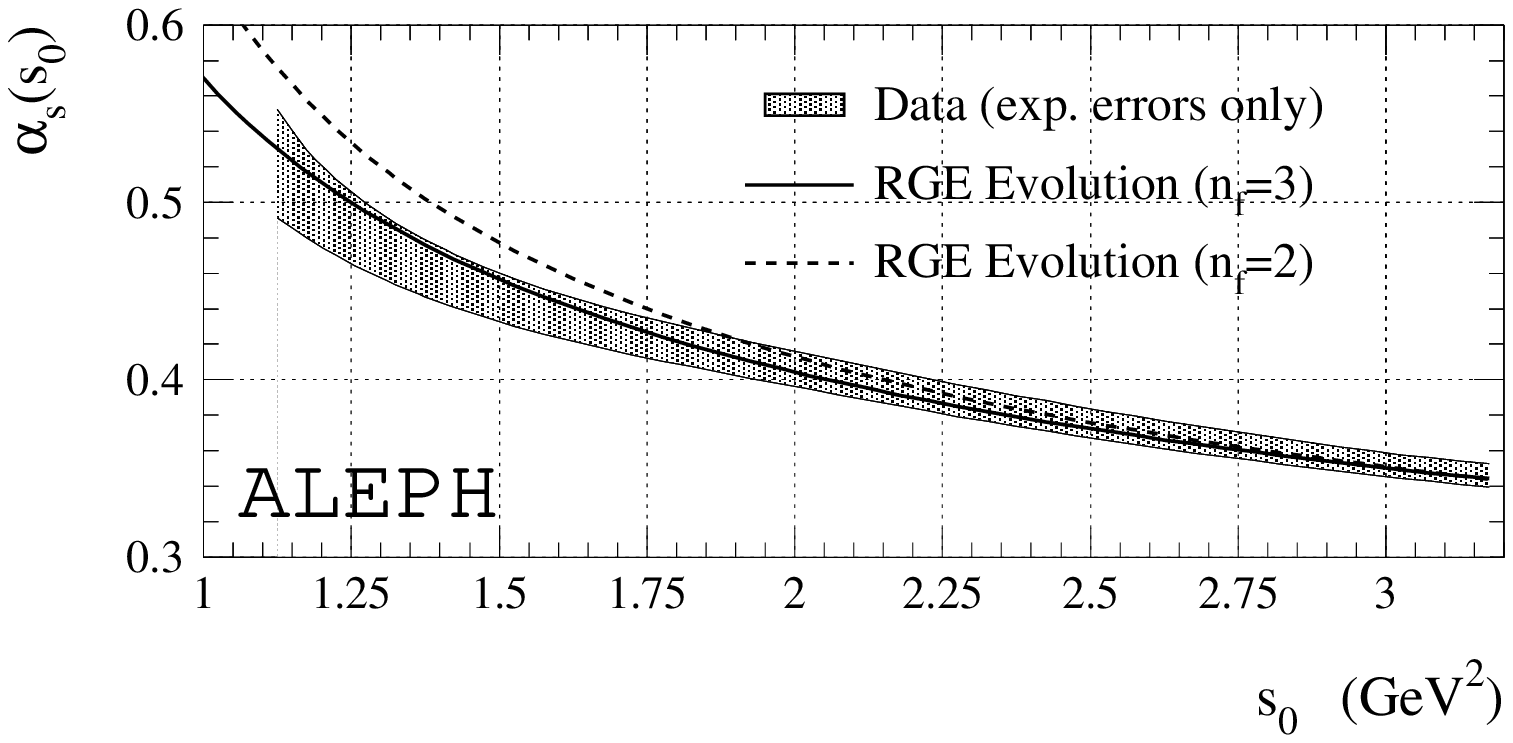,width=7.2cm}% 
        \psfig{file=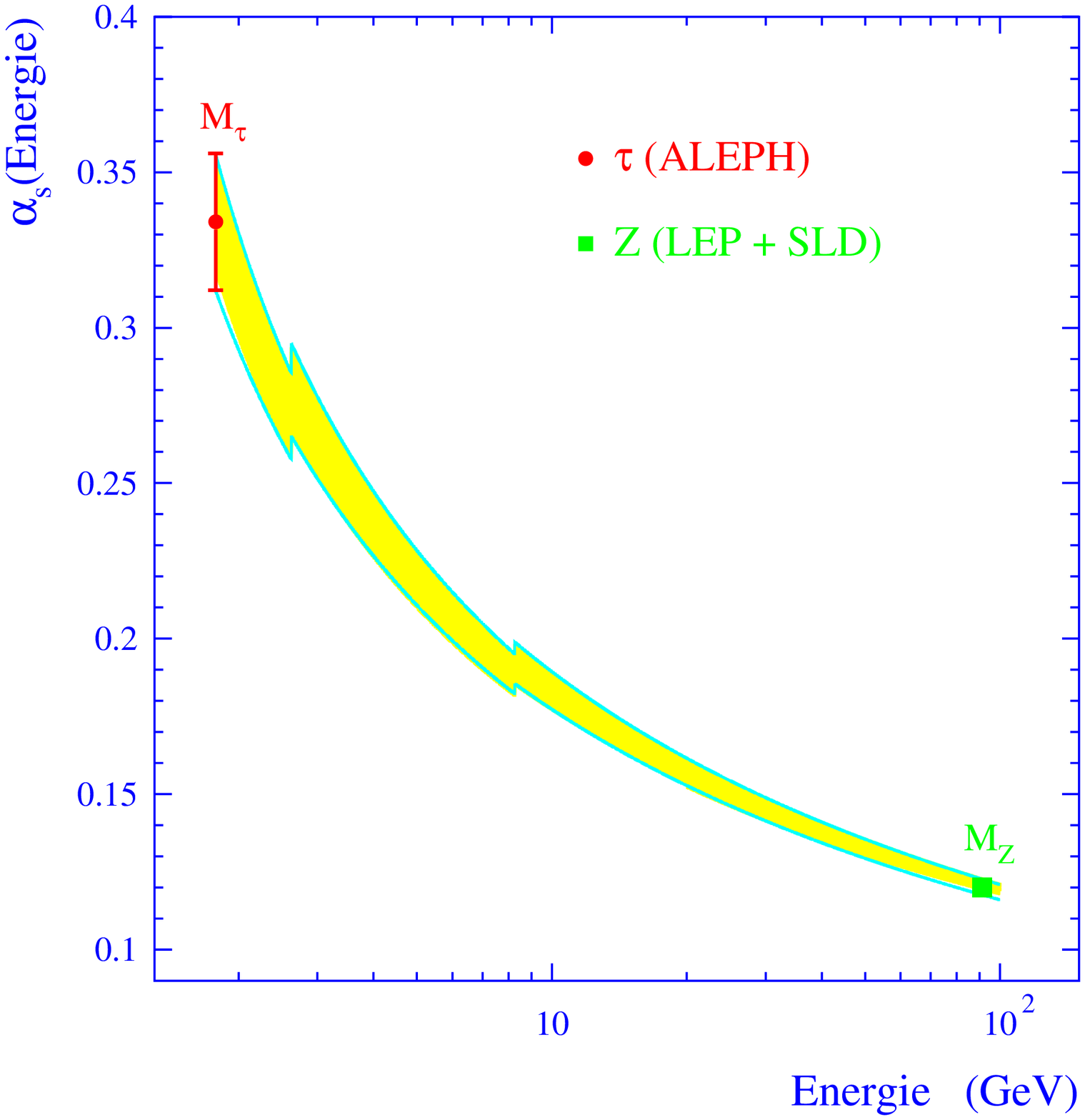,width=7cm}% 
        \caption{
          The running of $\alpha_s(s_0)$ obtained from the 
          fit of the theoretical prediction to \RtVpAs below
          the $\tau$ mass in the top plot: the shaded band shows the data 
          including experimental
          errors, while the curves give the four-loop RGE evolution 
          for two and three flavours. The plot below shows the
          evolution of the strong coupling (measured at $m_\tau^2)$
          to $M_Z^2$ predicted by QCD compared to the direct measurement.
          Flavour matching is accomplished at 3 loops at $2~m_c$ and $2~m_b$
          thresholds.}%
	\label{fig_runas}
\end{figure}
The result (\ref{alphaevol}) can be compared to the precise determination 
from the measurement of the $Z$ width, as obtained in the global 
electroweak fit. The variable $R_Z$ has similar advantages to $R_\tau$,
but it differs concerning the convergence of the perturbative
expansion because of the much larger scale. It turns out that this
determination is dominated by experimental errors with very small
theoretical uncertainties, \ie\ the reverse of the situation 
encountered in $\tau$ decays.
The most recent value~\cite{ewfit} yields 
$\alpha_s(M_{\rm Z}^2) = 0.1183 \pm 0.0027$, in excellent agreement
with (\ref{alphaevol}). 
Fig.~\ref{fig_runas} illustrates well the agreement between the evolution
of $\alpha_s(m_{\tau}^2)$ predicted by QCD and
$\alpha_s(M_{\rm Z}^2)$. 

\section{Strange Spectral Function and Strange Quark Mass}

The spectral function for strange final states has been determined by
ALEPH~\cite{aleph_ksum}: it is 
dominated by the vector $K^*(890)$ and higher mass (mostly axial-vector)
resonances.
The total rate for strange final states, using the complete ALEPH analyses 
supplemented by results from other experiments~\cite{dchpp} is determined to 
be $B(\tau \rightarrow \nu_\tau {\rm hadrons}_{S=-1})=(29.3\pm1.0)~10^{-3}$,
leading to
\beq
 R_{\tau,S}= 0.163 \pm 0.006.
\eeq

\begin{figure}
        \psfig{file=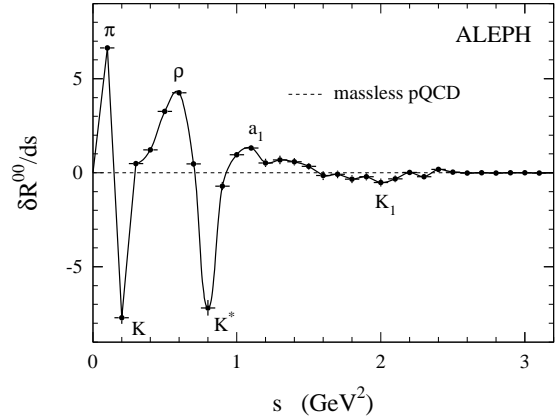,width=7.4cm}% 
        \caption{Differential rate for $\Delta_\tau^{00}$, difference between
       properly normalized nonstrange and strange spectral functions (see text
       for details). The contribution from massless perturbative QCD vanishes.
       To guide the eye, the solid line interpolates between bins of constant
       0.1 GeV$^2$ width.}%
	\label{diff_strange}
\end{figure}

Spectral moments are again useful tools to unravel the different components
of the inclusive rate. Since we are mostly interested in the specific
contributions from the $\overline{u}s$ strange final state, it is useful
to form the difference
\beq
 \Delta_\tau^{kl} \;\equiv\;
     \frac{1}{|V_{ud}|^2}R_{\tau,S=0}^{kl} - 
     \frac{1}{|V_{us}|^2}R_{\tau,S=-1}^{kl}
\eeq
where the flavour-independent perturbative part and gluon condensate cancel.
Fig.~\ref{diff_strange} shows the interesting behaviour of $\Delta_\tau^{00}$
expressed differentially as a function of $s$.
The leading QCD contribution to $\Delta_\tau^{kl}$ is a term proportional
to the square of the strange quark mass at the $\tau$ energy scale. 
We quote here the recent
result from the analysis of Ref.~\cite{cdghpp}, yielding
\beq
  m_s(m_\tau^2)~=~
       (120 \pm 11_{\rm exp} \pm 8_{V_{us}} \pm 19_{th})~{\rm MeV}\\
\eeq
where the dominant uncertainty is from theory, mostly because of the poor
convergence behaviour. 

\section{Conclusions}

The decays $\tau \rightarrow \nu_\tau$ + hadrons constitute a clean and
powerful way to study hadronic physics up to $\sqrt{s} \sim 1.8$ GeV.
Beautiful resonance analyses have already been done, providing new
insight into hadron dynamics. Probably the major surprise has been the 
fact that inclusive hadron production is well described by perturbative
QCD with very small nonperturbative components at the $\tau$ mass.
In spite of the fact that this low-energy region is dominated by resonance
physics, methods based on global quark-hadron duality work indeed very
well.

The ALEPH preliminary results using the full LEP1 sample have been 
presented. Satisfactory agreement with CLEO is observed in the $\pi \pi^0$
decay mode. The $\tau$ spectral functions
have now reached a precision level where detailed investigations are
possible, particularly in the most interesting $\pi \pi^0$ channel. The
breaking of SU(2) symmetry can be directly determined through the 
comparison between \ee\ and $\tau$ spectral functions. 
In particular, the possibility of a $\rho$ mass splitting between neutral
and charged states is carefully investigated. The data tend to favour 
non-degenerate states with $m_{\rho^-}-m_{\rho^0}=(2.3 \pm 0.8)$~MeV.
Correcting the $\tau$ spectral function for this additional isospin breaking
leaves an overall normalization difference of $3.3\%$ with \ee\ data, thus
enhancing the existing discrepancy for $a_\mu^{\rm had,LO}$. The situation
must be revisited with new high precision \ee\ data~\cite{kloe_pisa,md_pisa}.

The measurement of the vector and axial-vector spectral functions has
provided the way for quantitative QCD analyses. These spectral functions
are very well described in a global way by $O(\alpha_s^3)$ perturbative
QCD with small nonperturbative components. Precise determinations of
$\alpha_s$ agree for both spectral functions and they also agree with all
the other determinations from the Z width, the rate of Z to jets and
deep inelastic lepton scattering. Indeed from $\tau$ decays,
$\alpha_s(M_{\rm Z}^2)_\tau=0.1202 \pm 0.0027$,
in excellent agreement with the average from all other 
determinations~\cite{alphamor},
$\alpha_s(M_{\rm Z}^2)_{non-\tau}=0.1187 \pm 0.0020$.

The strange spectral function yields a value for the
strange quark mass which can be evolved to the usual comparison scale,
$m_s(2~{\rm GeV})=(116^{+20}_{-25})~{\rm MeV}$.

The $\tau$ spectral functions have been shown to be a privileged field
for the study of QCD.

\section*{Acknowledgments}

I would like to thank Shaomin Chen, Andreas H\"ocker,
Changzheng Yuan and Zhiqing Zhang for their many contributions 
to this work.  Congratulations to Marco Incagli and
Graziano Venanzoni for organizing a very stimulating workshop.

\end{document}